\documentclass[twocolumn,amsmath,amssymb]{revtex4}
\usepackage{graphicx}
\usepackage{bm}

\begin{document}
\title{2D Rashba system in AC magnetic field}
\author{I.I. Lyapilin}\email{Lyapilin@imp.uran.ru}
\author{A.E. Patrakov}
\affiliation{Institute of Metal Physics, UD of RAS, Yekaterinburg,
Russia}
\begin{abstract}
The response of an electron system to a DC measurement electric
field has been investigated in the case when the system is driven
out of the equilibrium by the magnetic ultra-high frequency field
that leads to combined transitions. The discussed model includes
contributions from Landau quantization and from microwave
irradiation. Impurity centers are considered as sources of
scattering. It has been shown that the perturbation of the
electron system by the ultra-high frequency magnetic field leads
to oscillations of the diagonal components of the conductivity
tensor.
\end{abstract}
\maketitle
\section{Introduction}

The interest in theoretical studies of transport phenomena in 2D electron
systems has raised substantially after the discovery \cite{Zudov03,Mani02}
of oscillations of the diagonal components of the conductivity tensor in
``ultraclean'' GaAs/Al$_{x}$Ga$_{1-x}$As samples in the classical interval
of the magnetic field intensity, where the Shubnikov---de Haas (SdH)
oscillations don't manifest themselves.
Together with the oscillations of the diagonal components of the
conductivity tensor caused by absorption by charge carriers of the energy of
the microwave radiation field and transitions between Landau levels,
``beats'' have been found experimentally \cite{Mani02} in the interval of
more weak magnetic fields.
Such beats are usually related with the manifestations of the interaction
between kinetic and spin degrees of freedom of the conductivity electrons.
Such interaction is the spin-orbit interaction (SOI) that is known to be the
origin of numerous effects in transport phenomena observed in such systems.
Among them there are, e.g., beats in SdH oscillations \cite{Das}, spin
accumulation \cite{Hammar-00}, magneto-electric effect \cite{Levitov}, etc.
SOI also leads to the possibility of electron transitions between Landau
levels in the magnetic field at the combined resonance frequencies
\cite{Rash}, thus transitions being possible both in antinodes of electric
and magnetic fields \cite{Kalash}.
Finally, operation of a spin transistor (schemes of which have been
considered in \cite{zutic}) is based upon spin degrees of freedom.
All of the above has determined elevated interest to investigations of the
SOI in 2D semiconductor structures.

For the purpose of studying the SOI, it appears interesting to investigate a
model in which the role of the SOI should manifest itself the strongest.
Since the SOI depends upon both translational and spin degrees of freedom,
then it is a channel over which energy (both electric and magnetic) can be
absorbed from the ultra-high frequency field, thus causing transitions
between Landau levels.
Because of that, it is interesting to investigate the response of a
non-equilibrium electron system to a DC weak (``measurement'') electric
field for the case when the initial non-equilibrium state is created with a
high-frequency AC magnetic field that leads to combined transitions.
The question is how this perturbation affects transport coefficients, in
particular, the conductivity tensor.

The discussed model includes the contributions fom Landau quantization and
(in th long-wavelength limit) from the microwave radiation exactly, without
use of the perturbation theory.
We consider impurity centers for the role of scatterers, treating the
scattering process perturbatively.

\section{Effective Hamiltonian}
The Hamiltonian of the system under consideration is:
\begin{equation}\label{1}
\mathcal H(t) = H_k+H_s+H_{ks}+H_{eh}(t)+H^0_{ef}+H_v+H_{ev}.
\end{equation}
Here $H-k$ and $H_s$ are kinetic and Zeeman energies, respectively, in the
magnetic field $\bm H = (0, 0, H)$:
\begin{multline}\label{2}
 H_k\,=\,\sum_i\frac{(\bm p_i-(e/c)\bm A (x_i))^2}{2m},\\
 H_s\,=\,\hbar\omega_s\,\sum\,S_i^z,\quad
 \hbar\omega_s=g\mu_0H.
\end{multline}
$S_i^\alpha$ and $p_i^\alpha$ are operators of the components of
the spin and kinetic momentum of the $i$th electron, where
$[p^\alpha,p_j^\beta] = -\delta_{ij} i m
\hbar\omega_c\varepsilon_{\alpha\beta z}$, $\omega_c=|e|H/mc$ is
the cyclotron frequency, and $\mu_0$ is Bohr magneton. $H^0_{ef}$
is the hamiltontal of the electrons' interaction with the electric
field $\bm E =(E_x, 0, 0)$:
\begin{equation}\label{02}
H_{ef} = -e\bm E \sum_i\bm r_i.
\end{equation}
$H_{eh}(t)$ is the interaction of electrons with the AC magnetic field:
\begin{equation}\label{002}
 H_{eh}(t) = g\mu_0 \bm H(t) \sum_i \bm S_i.
\end{equation}

$\bm H(t)= (H_x(t), H_y(t), H_z(t))$. $H_{ev}$ and $H_v$ are Hamiltonians of
the electron-lattice interaction and of the lattice itself, respectively.
$H_{ks}(p)$ is the interaction between translational and spin degrees of
freedom. Its most general form is:
\begin{equation}\label{3}
H_{ks}(\bm p) = \sum_j \bm f(\bm p_j) \bm S_j.
\end{equation}
Here $\bm f(\bm p_j)$ is a pseudo-vector, components of which can be
represented as a form of order $s$ in the components of the kinetic momentum
$p_j^\alpha$.

The spin-orbit interaction leads to correlation of spatial and spin motion
of electrons, thus, the translational and spin-related subsystems are not
well-defined.
Since the SOI is in some sense small, then one can perform a
momentum-dependent canonical transformation that decouples kinetic and spin
degrees of freedom.
Naturally, all other terms in the Hamiltonian, describing the interaction
of electrons with the lattice and external fields (if any) also undergo the
transformation.
In this case, the effective interaction of electrons in the system with
external fields appears, which leads to resonant absorption of the field
energy not only at the frequency of the paramagnetic resonance $\omega_s$
or cyclotron resonance $\omega_c$, but also at their linear combinations,
i.e. the combined resonance.
The gauge-invariant theory of the combined resonance has been developed in
\cite{VPKalash}.

Assuming the SOI to be small, we perform the canonical transformation of the
Hamiltonian.
Up to the terms linear in $T(t)$, we have:
\begin{equation}\label{5}
\tilde H = e^{T(p)}\mathcal H e^{-T(p)}\approx
\mathcal H +[T(p), \mathcal H].
\end{equation}
The operator of the canonical transformation $T(p)$ has to be determined
from the requirement that, after the transformation, the $k$ and $s$
subsystems become independent. This requirement can be written as the
following condition:
\begin{equation}\label{6}
H_{ks}(p) + [T(p), H_k + H_s]=0.
\end{equation}
Note that, after the canonical transformation, the $H_k$ and $H_s$ operators
are the integrals of motion if $H_{ev}=0$ and there is no interaction with
external fields.

We assume the specific form of the SOI term, namely, Rashba interaction,
which is non-zero even in the linear order in momentum:
\begin{equation}\label{7}
H_{ks}(p) =
\alpha \varepsilon_{zik} \sum_j S^i_j p^k_j =
\frac{i\alpha}{2} \sum_j (S^+_j p^-_j - S^-_j p^+_j),
\end{equation}
$$
S^\pm = S^x \pm i S^y, \quad p^\pm = p^x \pm i p^y.
$$
Here $\alpha$ is the constant characterizing the SOI, $\varepsilon$ is the
fully-antisymmetric Levi---Chivita tensor.

Now we find the explicit expression for the operator $T(p)$.
Inserting the operator (\ref{7}) into the general solution for Eq. (\ref{6})
and integrating, we obtain:
\begin{multline}\label{8}
T(p)=\frac{i}{\hbar}\lim_{\varepsilon\rightarrow
+0}\int\limits_{-\infty}^0dt\,e^{\varepsilon
t}\,e^{itH_o/\hbar}\,H_{ks}(p)\,e^{-itH_o/\hbar}=\\
=\frac{i\,\alpha}{2\hbar(\omega_o-\omega_s)}
\sum_j\,(S^+_jp^-_j-S^-_jp^+_j).
\end{multline}
Obviously, the criteria for the applicability of this theory is
that, for characteristic values of the electron momentum $\bar p$,
the inequality $\alpha \bar p \ll \hbar(\omega_c-\omega_s)$ should
hold.

One can write the transformed Hamiltonian in the following form:
\begin{multline}\label{9}
\mathcal{\tilde{H}}(t)\,=\,H_0 + H^0_{ef}
+H_{eh}(t)+[T(p),\,H_{eh}(t)+\\+H^0_{ef}+{H}_{ev}].
\end{multline}
$$
H_0 = H_k + H_s + H_v + H_{ev}.
$$
Using the explicit expression for the operator $T(t)$, we find:
\begin{multline}\label{10}
g\mu_0[T(p),S^\alpha]H^\alpha(t)\,=
\frac{i\,g\,\alpha\,\mu_0}{2\hbar(\omega_c-\omega_s)}
\,\{(\,T^{z-}H^+(t)-\\-T^{z+}\,H^-(t)\,)+ (T^{-+}-T^{+-})H^z(t)\}.
\end{multline}
$$
T^{\alpha \beta}=\sum_i S^\alpha_i p^\beta_i.
$$
Now we find the operators of power $\dot
H_{i(h)}(t)=(i\hbar)^{-1}[H_i,H_{eh}(t)+[T(p),H_{eh}(t)]]$
absorbed by kinetic ($i=k$) and spin ($i=s$) subsystems due to the
interaction of electrons with the AC magnetic field. We have:
\begin{multline}\label{18}
\dot{H}_{k(h)}(t)=
\frac{g\alpha\mu_0\omega_c}{2\hbar(\omega_c-\omega_s)}
\,\{(T^{-+}+T^{+-})H^z(t)-\\-(T^{z-}H^+(t)+T^{z+}H^-(t))\},
\end{multline}
\begin{multline}
\dot{H}_{s(h)}(t)=\frac{i\omega_sg\mu_0}{2}
(\,S^-H^+(t)-S^+H^-(t))-\\-\frac{g\alpha\mu_0\omega_c}
{2\hbar(\omega_c-\omega_s)}(T^{-+}+T^{+-})H^z(t)
\end{multline}
The total absorbed power can be written as:
\begin{equation}\label{19}
\dot{H}_{k(h)}(t)+\dot{H}_{s(h)}(t)= J_m^\beta\,H^\beta(t),
\end{equation}
$$
J_m^\beta=\frac{g\mu_0}{i\hbar}
[S^\beta,[T(p),S^\beta],H_k(p)+H_s].
$$

The interaction of the spin degrees of freedom of the conductivity electrons
with the AC magnetic field $H_{eh}(t)$ leads to resonant transitions at the
frequency $\omega_s$.
However, as one can see from the expressions above, the effective
interaction $[T(p), H_{eh}(t)]$ leads to combined transitions at frequencies
$\omega_c \pm \omega_s$ and the cyclotron frequency $\omega_c$.
Since, for our further calculations, the response of the non-equilibrium
system to the measurement electric field is interesting, in which the
contribution from the translational degrees of freedom dominates, we will
restrict our consideration to the effective interaction solely.
Besides that, we limit the consideration to the case when the DC and AC
magnetic fields are perpendicular to each other:
$\bm H(t)= (H_x(t),H_y(t),0)$.
In this case, the effective interaction responsible for the combined
transitions has the following form:
\begin{multline}\label{20}
H_{eh,1}(t)=[T(p),\,H_{eh}(t)]= \\=\frac{i \alpha \omega_{1
s}}{2\hbar(\omega_c-\omega_s)} \,\sum_j S_j^z (\,p_j^+ e^{-i\omega
t}-p_j^- e^{i\omega t}\,).
\end{multline}
$\omega_{1s}=g e H_1/(2 m_0 c)$.  $H_1$ is the intensity of the
circularly polarized magnetic field, rotating with the frequency
$\omega$.

The dependence of the effective interaction $H_{eh,1}(t)$ upon time causes
certain difficulties while calculating the non-equilibrium response of the
electron system to the measurement electric field.
Thus, it is expedient to carry out one more canonical transformation
(Appendix A), that
removes the interaction $H_{eh,1}(t)$ and renormalizes the electron-impurity
interaction Hamiltonian (Appendix B), which acquires the time dependence then.
In the canonically transformed system, impurities act as a coherent
oscillating field that leads to resonant transitions.

\section{Non-Equilibrium Response}

We assume that the initial non-equilibrium state of the system under
consideration is created by the ultra-high frequency magnetic field and can
be described with the distribution $\bar{\rho}(t)$.
Obviously, if some
additional perturbation acts upon the system, then a new non-equilibrium
state is formed in the system, that requires an extended set of basis
operators for its description.
The new non-equilibrium distribution is described with the operator
$\rho(t,0)$.
The task is to find the response of a non-equilibrium system to a weak
measurement field.

We write the operator $\rho(t)$ using the integral representation for the
non-equilibrium statistical operator \cite{VPK}. In the linear approximation
in the external field $\bm E$, we have:
\begin{multline}\label{g38}
   \rho(t)=\bar{\rho}(t)-i \int\limits_{-\infty}^0\,dt_1\,
   e^{\varepsilon\,t_1}\,U(t+t_1)\times\\\times iL_{ef}^0
   \rho(t+t_1)\,U^+(t+t_1).
  \end{multline}
$$
i L_{ef}^0 A = \frac{1}{i\hbar} [A, H^0_{ef}].
$$
The operator $\bar{\rho}(t)$ satisfies the equation
\begin{equation}\label{g31}
\frac{\partial\bar{\rho}(t)}{\partial t} + \frac{1}{i\hbar}
[\bar{\rho}(t), H(t)] =-\varepsilon (\bar{\rho}(t) - \rho_q(t)),
\end{equation}
which has the following solution:
\begin{multline}\label{g32}
\bar{\rho}(t)\,=\,\rho_q(t)-\int\limits_{-\infty}^0\,dt_1\,
e^{\varepsilon\,t_1}\,U(t+t_1)\times\\\times\{\frac{\partial}{\partial
t}+iL(t)\} \rho_q(t+t_1)\,U^+(t+t_1),
\end{multline}
$$
i L(t) A = \frac{1}{i\hbar} [A, H(t)],
$$
where $U(t_1)$ is the evolution operator that is in fact a chronologically
ordeted exponent and satisfies the differential equation
$$
\frac{\partial U(t)}{\partial t} = \frac{1}{i\hbar} H(t) U(t).
$$
%\begin{equation}\label{g33}
%U(t) =\exp_+\{\frac{1}{i\hbar}\int\limits_{-\infty}^t e^{\varepsilon\tau}
% H(\tau) d\tau\}.
%\end{equation}
The quasiequilibrium statistical operator $\rho_q(t)$ can be expressed as:
\begin{equation}\label{g34}
\rho_q(t) = e^{-S(t)},\qquad S(t) = \Phi(t) + \sum_n P^+_n F_n(t),
\end{equation}
where $S(t)$ is the entropy operator, $\Phi$ is the Massieu---Planck
functional. $P_n$, $F_n$ are sets of the basis operators and their conjugate
functions, that describe the electron system.

Characterizing the state of the system by the mean values of the
operators $H_k$, $\bm p$, $H_s$, $N$, $H_v$ ($N$ being the
operator of the number of electrons), we obtain for the entropy
operator:
\begin{multline}\label{g35}
S(t)\,=\,\Phi(t)\,+\,\beta_k\,(H_k-\bm V(t)\,\bm p -\mu'N)+\beta_s
H_s\,+\\+ \beta\,H_v= S^0+\Delta S(t).
\end{multline}
$$
\Delta S(t)= -\beta_k \bm V(t) \bm p.
$$
Here $\beta_k$, $\beta_s$, $\mu' = \mu - mV^2/2$, $\bm V$, $\beta$
are parameters thermodynamically conjugate to the average values
of the introduced operators. They have the meaning of inverse
effective temperatures of the kinetic and spin electron
subsystems, chemical potential, drift velocity and the inverse
lattice temperature. Introduction of the effective temperatures
allows one to treat effects related to ``heating'' of the electron
and spin subsystems by external fields.

A linear addmittance corresponding to an arbitrary operator $B$ in the case
when the external force oscillates harmonically in time with the frequency
$\omega_1$ can be expressed as:
\begin{multline}\label{g36}
\chi_{BA}(t,\omega_1)=-\int\limits_{-\infty}^0 dt_1
e^{(\varepsilon-i\omega_1)t_1} \frac{1}{i\hbar}\operatorname {Sp}
\{ B\times\\\times e^{it_1L}[A, \bar{\rho}(t+t_1,0)] \}
\end{multline}

Within the framework described above, the task of the admittance
calculation is reduced to obtaining the transport matrix
$T_{BA}(t,\omega_1)$, which plays in the non-equilibrium case the
same role as in the case of the equilibrium response:
\begin{equation}\label{g37}
\chi_{BA}(t,\omega_1)=\chi_{BA}(t,0)
\frac{T_{BA}(t,\omega_1)+\varepsilon}{T_{BA}
(t,\omega_1)+\varepsilon-i\omega_1}.
\end{equation}
\begin{equation}\label{g38}
\chi_{BA}(t,0)=\langle B,A \rangle,\qquad T_{BA}=\frac{1}{\langle
B,A \rangle^{\omega_1}} \langle B,\dot{A}\rangle^{\omega_1}.
\end{equation}
Here
\begin{equation}\label{g39}
\langle B,A \rangle=-\frac{1}{i\hbar} \int\limits_{-\infty}^0 dt_1
e^{\varepsilon t_1} \operatorname{Sp} \{ B e^{it_1L} [A,\bar{\rho}(t+t_1,0)]
\},
\end{equation}
\begin{multline}\label{g40}
\langle B,A \rangle^{\omega_1}=-\frac{1}{i\hbar}
\int\limits_{-\infty}^0 dt_1 e^{(\varepsilon -i\omega_1)t_1}
 dt_2 e^{\varepsilon t_2} \operatorname{Sp} \{ B\times\\\times
e^{i(t_1+t_2)L} [A, \bar{\rho}(t+t_1+t_2,0)] \}.
\end{multline}

The real part of the transport matrix determines the relaxation frequency of
the non-equilibrium electrons' momentum.

\section{Momentum Relaxation Rate}

Assuming the temperatures of the translational and spin subsystems to be
equal (that corresponds to neglecting any heating effects), in the Born
approximation upon the electron-scatterer interaction we obtain for the
relaxation frequency:
\begin{multline}\label{t1}
\frac{1}{\tau} = \frac{\beta}{2 m n} \operatorname{Re}
\frac{1}{i\hbar} \int\limits_{-\infty}^0 dt_1 e^{(\varepsilon -
\omega_1)t_1}  dt_2 e^{\varepsilon t_2} \int\limits_0^1 d\lambda
\times{} \\ {}\times \operatorname{Sp}\{ \dot p^+_{(\tilde v)}(t)
e^{i L_0 (t_1 + t_2)}\rho_q^{\lambda} [\dot p^-_{(\tilde
v)}(t+t_1+\\+t_2), H_k + H_s] \rho_q^{1-\lambda}\}, \quad
p^\alpha_{(\tilde v)} = \frac{1}{i \hbar}[p^\alpha, \tilde
H_{ev}].
\end{multline}
Now we expand the formula (\ref{t1}) using the explicit expression
for the renormalized electron-impurity interaction
$\tilde{H}_{ev}$ from Appendix B.

It is convenient to introduce the following notation:
\begin{multline}\label{t2}
A(\lambda, t_1 + t_2) = \operatorname{Sp}\{ \dot p^+_{(\tilde
v)}(t) e^{i L_0 (t_1 + t_2)}\times\\\times\rho_q^{\lambda}[\dot
p^-_{(\tilde v)}(t+t_1+t_2), H_k + H_s] \rho_q^{ 1-\lambda}.
\end{multline}
Inserting the explicit expression for the electron-impurity
interaction and averaging over the system of scatterers, we
obtain:
\begin{multline}\label{t3}
A(\lambda, t_1 + t_2) =  \sum\limits_{\bm q j j' l} |V(\bm q)|^2
N_i e^{-i l \omega(t_1 + t_2)} J_l^2(|K_q|)
q^2\times{}\\{}\times\operatorname{Sp}\{ 2 S_j^z e^{i \bm q \bm
r_j} e^{i L_0(t_1 + t_2)} \rho_q^{\lambda} [ 2 S_{j'}^z e^{-i \bm
q \bm r_{j'}}, H_k + H_s] \rho_q^{1-\lambda}\}.
\end{multline}
Here $N_i$ is the impurity concentration. Rewriting this
expression using secondary quantization and averaging the electron
operators using Wick's theorem, we have:
\begin{multline}\label{t4}
A(\lambda, t_1 + t_2) = \sum\limits_{\bm q \nu \mu l} |V(\bm q)|^2
N_i e^{-i l \omega(t_1 + t_2)}\times\\\times
e^{-\frac{i}{\hbar}(t_1 + t_2)(\varepsilon_\mu -
\varepsilon_{\mu'})}
e^{-\beta_e(\varepsilon_\mu - \varepsilon_{\mu'})\lambda}\times{} \\
{}\times J_l^2(|K_q|) q^2 (\varepsilon_{\nu} - \varepsilon_{\mu})
|(2 S^z e^{i \bm q \bm r})_{\nu \mu}|^2 f_\nu (1 - f_\mu),
\end{multline}
where $f$ is the Fermi---Dirac distribution.

After integration over $\lambda$, $t_1$ and $t_2$, this produces:
\begin{multline}\label{t5}
\int\limits_{-\infty}^0 dt_1 e^{(\varepsilon - i \omega_1) t_1}
 dt_2 e^{\varepsilon t_2} \int\limits_0^1
d\lambda A(\lambda, t_1 + t_2) = \\=\sum\limits_{\bm q \nu \mu l}
|V(\bm q)|^2 N_i J_l^2(|K_q|) q^2|(2 S^z e^{i \bm q \bm r})_{\nu
\mu}|^2 (f_\nu - f_\mu)\times\\\times \frac{(\varepsilon - i l
\omega - (i/\hbar)(\varepsilon_\mu - \varepsilon_\nu))^{-1}}
{\varepsilon - i\omega_1 - i l \omega - (i/\hbar)(\varepsilon_\mu
- \varepsilon_\nu)}
\end{multline}
In the $\varepsilon \to 0$ limit, we obtain:
\begin{multline}\label{t6}
\frac{(i\hbar)^{-1}}{\varepsilon - i l \omega -
(i/\hbar)(\varepsilon_\mu - \varepsilon_\nu)} = \\= \left(
\operatorname{P}\frac{1}{l\hbar\omega +(\varepsilon_\mu -
\varepsilon_\nu)} - i \pi \delta(l\hbar\omega +(\varepsilon_\mu -
\varepsilon_\nu)) \right),
\end{multline}
where $\operatorname{P}$ denotes the Cauchy principal value.
Taking the $\omega_1 \to 0$ limit (because calculating a
zero-frequency response is our goal), we have:
\begin{multline}\label{t7}
\Delta(\frac{1}{\tau}) = -\frac{\pi\hbar}{2 m n} \sum\limits_{\bm
q \mu \nu l} \int d\mathcal E |V(q)|^2 N_i J_l^2(|K_q|) q^2
\times{} \\ {} \times |(2 S^z e^{i \bm q \bm r})_{\nu\mu}|^2
(f(\mathcal E + l \hbar\omega) - f(\mathcal E))\times\\\times
\delta(\mathcal E - \varepsilon_\mu)
\frac{\partial}{\partial\mathcal E} \delta(l\hbar\omega + \mathcal
E - \varepsilon_\nu)
\end{multline}

The equation (\ref{t7}) contains a singularity in its right hand side, which
is removed, as usual, due to broadening of the Landau levels by scattering
electrons on impurities:
\begin{equation}\label{t8}
\delta(\mathcal E - \varepsilon_\mu) \to D_\mu(\mathcal E)=
\frac{\sqrt{\pi/2}}{\Gamma} \exp\left(-\frac{(\mathcal E -
\varepsilon_\mu)}{2 \Gamma^2}\right).
\end{equation}
The Landau level width $\Gamma$ can be expressed via the electron mobility
$\mu$ in zero magnetic field:
\begin{equation}\label{t9}
\Gamma = \hbar\sqrt\frac{2 \gamma_n \omega_c}{\pi \tau_{tr}}, \qquad
\tau_{tr}=\frac{m \mu}{|e|}
\end{equation}

Note that, if $T > \Gamma$, one can pull $f(\mathcal E \pm
\hbar\omega) - f(\mathcal E)$ out of the $\mathcal E$ integral as a
slowly-changing factor. This yields:
\begin{multline}\label{t10}
\int d\mathcal E \frac{\partial}{\partial \mathcal E}
D_\nu(\mathcal E \pm \hbar\omega) D_\mu(\mathcal E) =\\
=-\frac{\pi^{3/2}(\varepsilon_\mu - \varepsilon_\nu \pm
\hbar\omega)}{4\Gamma^3} \exp\left(-\frac{(\varepsilon_\mu -
\varepsilon_\nu \pm \hbar\omega)^2} {4 \Gamma^2}\right)
\end{multline}

The wave functions upon which the matrix elements in (\ref{t7}) are
calculated, have the form:
\begin{multline}\label{t11}
\psi_{\nu}\equiv\psi_{n k^x S^z} = \frac{1}{\sqrt{2^n n! \pi^{1/2}
\ell}} \exp(i
k^x x)\times\\
\times\exp\left(-\frac{(y - y_0)^2}{2 \ell^2}\right) H_n(\frac{y -
y_0}{\ell})\chi_{S^z}
\end{multline}

$y_0=\ell^2 k^x$ is the cyclotron orbit center coordinate, $\ell$
is the magnetic length, $H_n(x)$ denotes Hermite polynomials, and
$\chi_{S^z}$ is the eigenfunction of the $z$ spin projection.
Calculation of the matrix element in (\ref{t7}) produces:
\begin{multline}\label{t12}
|\langle \nu | 2 S^z \exp(i \bm q \bm r) | \mu \rangle|^2 =
\exp\left(-\frac{\ell^2 q^2}{2}\right) \times{}\\{}\times
\frac{(\min(n_\nu, n_\mu))!}{(\max(n_\nu, n_\mu))!}
\left(\frac{\ell^2 q^2}{2}\right)^{|n_\nu - n_\mu|}\times\\
\times\left(L_{\min(n_\nu, n_\mu)}^{|n_\nu - n_\mu|}\left(
\frac{\ell^2 q^2}{2}\right)\right)^2 \delta_{S^z_\nu,S^z_\mu}
\delta_{k^x_\nu, q_x + k^x_\mu}
\end{multline}
In the case of sufficiently weak AC magnetic field one can neglect
terms with $|l|>1$ and use the approximation $J_{\pm 1}(x) = \pm
x/2$. As a result of such simplifications, we obtain:
\begin{multline}\label{t13}
\Delta(\frac{1}{\tau}) = \frac{\hbar}{8 m n \ell^2}\sum\limits_{
\bm q n_\nu n_\mu l=\pm 1 } |V(q)|^2 N_i |K_q|^2 q^2
\times\\
\times\exp\left( -\frac{\ell^2 q^2}{2}\right) \frac{(\min(n_\nu,
n_\mu))!}{(\max(n_\nu, n_\mu))!} \left(\frac{\ell^2
q^2}{2}\right)^{|n_\nu - n_\mu|}\times{}\\{}\times
\left(L_{\min(n_\nu, n_\mu)}^{|n_\nu - n_\mu|}(\frac{\ell^2
q^2}{2})\right)^2 (f(\varepsilon_\nu) - f(\varepsilon_\mu))\times\\
\times\frac{\pi^{3/2}(\varepsilon_\mu - \varepsilon_\nu + l
\hbar\omega)}{4\Gamma^3} \exp\left(-\frac{(\varepsilon_\mu -
\varepsilon_\nu + l \hbar\omega)^2} {4 \Gamma^2}\right)
\end{multline}

Integrating over $\bm q$, we have:
\begin{multline}\label{t14}
\int\limits_0^\infty d(q^2) q^4 \exp\left(-\frac{\ell^2
q^2}{2}\right) \left(-\frac{\ell^2 q^2}{2}\right)^{|n_\nu -
n_\mu|}\times\\\times \left(L_{\min(n_\nu, n_\mu)}^{|n_\nu -
n_\mu|}(\frac{\ell^2 q^2}{2})\right)^2 =
\frac{8}{\ell^6}\frac{(\max(n_\nu, n_\mu))!}{(\min(n_\nu,
n_\mu))!}\times\\\times (n_\nu^2 + n_\mu^2 +3 (n_\nu + n_\mu) + 4
n_\nu n_\mu + 2)
\end{multline}

Thus, for the case of point scatterers, where $V(q)$ does not depend on
$q$, the radiation-induced correction to the inverse relaxation time is:
\begin{multline}\label{t15}
\Delta(\frac{1}{\tau}) = \frac{\hbar}{4 m n \ell^8}\sum\limits_{
n_\nu n_\mu l=\pm 1 }\frac{|V(q)|^2 N_i\alpha^2
\omega_{1s}^2}{(\omega_c - \omega_s)^2(\omega -
\omega_c)^2}\times\\
\times (n_\nu^2 + n_\mu^2 + 3 (n_\nu + n_\mu) + 4 n_\nu n_\mu + 2)
\times{}\\{}\times (f(\varepsilon_\nu)-f(\varepsilon_\mu))
\frac{\pi^{1/2}(\varepsilon_\mu - \varepsilon_\nu + l
\hbar\omega)}{\Gamma^3} \times\\
\times\exp\left(-\frac{(\varepsilon_\mu - \varepsilon_\nu + l
\hbar\omega)^2} {4\Gamma^2}\right)
\end{multline}
Using the expression for the momentum relaxation rate, one can also write
the formula for the diagonal componetns of the conductivity tensor
$\sigma_{x x}$, according to which the numerical calculations in the next
section are carried out:
\begin{equation}\label{t16}
\sigma_{x x} =\frac{n
e^2}{m}\frac{\tau^{-1}}{\omega_c^2+\tau^{-2}}.
\end{equation}

\section{Numerical Analysis}

Numerical calculations of the diagonal components of the
conductivity tensor according to Eq. (\ref{t16}) have been carried
out with the following parameters: $m = 0.067 m_0$ ($m_0$ is the
free electron mass), the Fermi energy is $\mathcal{E}_F = 10$~meV,
the mobility of the 2D electrons varies as $\mu\approx 0.1 -
1.0\times 10^7$~cm$^2$/Vs, the electron density $n=3\times
10^{11}$~cm$^{-2}$. The microwave radiation frequency is $f=
50$~GHz, the temperature is $T\approx 2.4$~K. The magnetic field
varied as 0.02~--~0.3~T.

\begin{figure}[t]
\center\includegraphics[width=8cm]{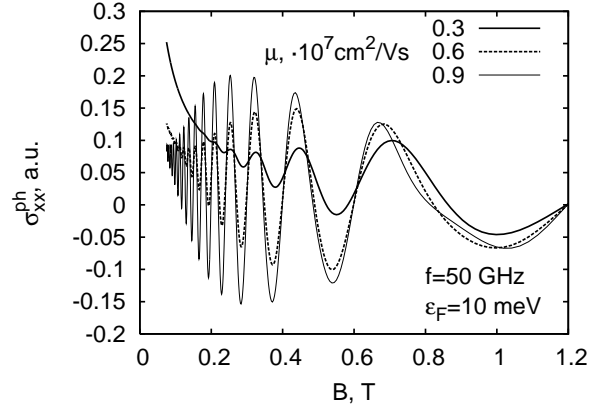}
\caption{Photoconductivity of the 2D electron gas vs of the
magnetic field induction for different values of electron
mobility. The radiation frequency is 50~GHz and
$\gamma=2$.}\label{fig:Lmuhmu}
\end{figure}

The dependence of the 2D electron gas photoconductivity on the
$\omega/\omega_c$ ration is presented in Fig.~\ref{fig:Lmuhmu}.
One can see that the dependence of electron mobility upon the
magnetic field has the oscillating character. In the region of low
magnetic field, the oscilation amplitude drops significantly when
zero-magnetic-field mobility is decreased.

\begin{figure}[t]
\center\includegraphics[width=8cm]{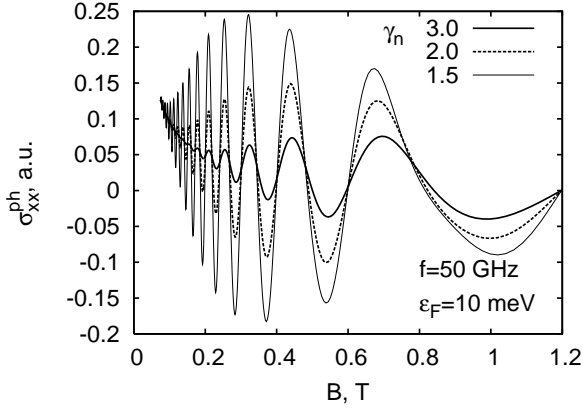}
\caption{Photoconductivity of the 2D electron gas vs of the
magnetic field induction for different values of electron
mobility. The radiation frequency is 50~GHz and
$\gamma=2$.}\label{fig:LmuHG}
\end{figure}

In Fig.~\ref{fig:LmuHG}, the photoconductivity dependence upon the
magnetic field is presented for different values of $\gamma_n$ and the same
microwave radiation frequency 50 GHz and electron mobility  $\mu=
0.6\times 10^7$~cm$^2$/Vs.
 As one can see, the oscillation amplitude is very
sensitive to the width of Landau levels.

\section{Conclusion}
The response of a non-equilibrium electron system to the DC electric
measurement field has been studied for the case when the initial
non-equilibrium state of the system is created by an ultra-high frequency
magnetic field that leads to combined transitions.
Within the proposed theory, it has been shown that such
perturbation of the electron system essentially
influences the transport coefficients and leads to the oscillations of the
diagonal components of the conductivity tensor.
The discussed effect is analogous to the phenomenon observed in
GaAs/AlGaAs heterostructures with ultra-high electron mobility \cite{Mani02}.
However, unlike that phenomenon, the manifestation of the oscillatory
pattern is dictated by the spin-orbit interaction existing in the crystals
under consideration.

\appendix
\section{}

In this Appendix, a canonical transformation $W_2(t)$ is built, that
excludes the renormalized interaction with the AC magnetic field from the
effective Hamiltonian.
The canonical transformation operator is searched from the equation:
\begin{multline}i\label{eq:UC2}
W_2^\dagger(t) (-i\hbar \frac{\partial}{\partial t} +H_k + H_s +
H_{eh,1}(t))W_2(t) =\\
= -i\hbar \frac{\partial}{\partial t} +H_k +H_s.
\end{multline}
The operator $W_2(t)$ is searched in the following form:
\begin{multline}
W_2(t) =  \exp(i \sum\limits_j(\eta^-(t)p_j^+ S_j^z +
\eta^+(t)p_j^- S_j^z))\times\\\times\exp(i\theta(t)),
\end{multline}
where one has to determine the parameters $\theta(t)$ and
$\eta^\pm(t)$.

In order to determine these parameters, the canonical transformation is
applied to all terms in the left-hand side of Eq. (\ref{eq:UC2}):
\begin{multline}
W_2^\dagger(t) (-i\hbar \frac{\partial}{\partial t}) W_2(t) =
-i\hbar \frac{\partial}{\partial t} + \hbar\dot\theta(t)+\\
+\hbar\sum\limits_j(\dot\eta^-(t)p_j^+ + \dot\eta^+(t)p_j^-)
\end{multline}
\begin{multline}
W_2^\dagger(t) H_k W_2(t) = \frac{1}{4m}\sum\limits_j (p_j^+
p_j^-\\
 -  4 i m \hbar \omega_c \eta^+(t) S_j^z p_j^- +
4 i m \hbar \omega_c \eta^-(t) S_j^z p_j^+\\ + m^2 \hbar^2
\omega_c^2 \eta^+(t) \eta^-(t))
\end{multline}
\begin{equation}
W_2^\dagger(t) H_s W_2(t) = H_s
\end{equation}
\begin{multline}
W_2^\dagger(t) H_{eh,1}(t) W_2(t) =
\frac{i\alpha\omega_{1s}}{2(\omega_c - \omega_s)}\sum\limits_j
S_j^z( (p_j^+\\ - 2 i m \hbar\omega_c \eta^+(t)S_j^z)e^{-i\omega
t} -{}\\{}- (p_j^- + 2 i m \hbar\omega_c \eta^-(t)S_j^z)e^{i\omega
t} )
\end{multline}

Obviously, the equation (\ref{eq:UC2}) becomes an identity under the
following conditions:
\begin{equation}
\hbar\dot\eta^\pm(t) \mp i \hbar \omega_c \eta^\pm(t) \mp \frac{i
\alpha \omega_{1s}}{2(\omega_c - \omega_s)}e^{\pm i \omega t} = 0,
\end{equation}
\begin{multline}
\hbar\dot\theta(t) + \frac{N m \hbar^2 \omega_c^2}{4} \eta^+(t)
\eta^-(t) +\\+\frac{N \alpha \omega_{1s} m \hbar
\omega_c}{4(\omega_c - \omega_s)} (\eta^+(t) e^{-i \omega t} +
\eta^-(t) e^{i\omega t}) = 0.
\end{multline}

Thus one can write down the explicit expressions for $\eta^\pm(t)$,
$\theta(t)$:
\begin{equation}
\eta^\pm(t) = \frac{\alpha\omega_{1s}}{2\hbar(\omega_c -
\omega_s)(\omega - \omega_c)} e^{\pm i \omega t},
\end{equation}
\begin{equation}
\theta(t) = \frac{N m \alpha^2 \omega_{1s}^2 (3\omega_c - 4\omega)t}
{16 \hbar(\omega_c - \omega_s)^2 (\omega - \omega_c)^2}
\end{equation}

Therefore, the explicit form of the canonical transformation $W_2$ is known.
The physical meaning of this transformation is the change to two
non-uniformly translationally moving reference frames, different for
electrons with opposite spin directions.

\section{}

Since we consider only elastic scattering, the electron-impurity
interaction Hamiltonian has the form:
\begin{equation}
H_{ev}= \sum\limits_{\bm q j} V(\bm q) \rho(\bm q)e^{i \bm q \bm
r_j},
\end{equation}
where $V(\bm q)$ is a Fourier component of the potential created
by a single impurity corresponding to the wave vector $\bm q$,
$\rho(\bm q)$ is a Fourier component of the impurity number
density.

As a result of the canonical transformation $W_2(t)$,
renormalization of the electron-impurity interaction happens. For
obtaining the renormalized Hamiltonian of the electron-impurity
interaction, it is sufficient to calculate $W_2^\dagger(t) \exp(i
\bm q \bm r_j) W_2(t)$. Using the explicit form of the operator
$W_2^\dagger(t)$, we obtain:
\begin{equation}\label{eq:exp-to-expand}
W_2^\dagger(t) e^{i \bm q \bm r_j} W_2(t) = \exp(i \bm q (\bm
r_j+\bm{\Delta r}_j)),
\end{equation}
where
\begin{equation}\label{eq:deltar:bad}
\bm{\Delta r}_j = -\frac{\alpha\omega_{1s}}{(\omega_c -
\omega_s)(\omega -\omega_c)}S_j^z\times (\cos\omega t, \sin\omega
t, 0)
\end{equation}

Expansion of the exponent in (\ref{eq:exp-to-expand}) in terms of
Bessel functions $J_l(x)$ yields:
\begin{equation}
e^{-i \operatorname{Re}(2 K_q S_j^z e^{i\omega t})} =
\sum\limits_{l=-\infty}^{\infty} \left(2 S_j^z
\frac{K_q}{i|K_q|}e^{i\omega t}\right)^l J_l(|K_q|),
\end{equation}
$$
K_q = \frac{\alpha \omega_{1 s} ( q_x - i q_y )}{( \omega_c
  - \omega_s ) ( \omega - \omega_c )}.
$$
Thus, the renormalized Hamiltonian of the electron-impurity
interaction is:
\begin{multline}
\tilde H_{ev}(t)= \sum\limits_{\bm q
j}\sum\limits_{l=-\infty}^{\infty} V(\bm q) \rho(\bm q)e^{i \bm q
\bm r_j}\times\\\times
 \left(2 S_j^z \frac{K_q}{i|K_q|}\,e^{i\omega
t}\right)^l J_l(|K_q|).
\end{multline}

\end{document}